\documentclass[a4paper,twocolumn,pra]{revtex4-1}
\usepackage{amsmath}
\usepackage{amssymb}
\usepackage{graphicx}
\usepackage{amsthm}
\usepackage{hyperref}
\usepackage{bbold}

\begin{document}

\newcommand*{\cl}[1]{{\mathcal{#1}}}
\newcommand*{\bb}[1]{{\mathbb{#1}}}
\newcommand{\ket}[1]{|#1\rangle}
\newcommand{\bra}[1]{\langle#1|}
\newcommand{\inn}[2]{\langle#1|#2\rangle}
\newcommand{\proj}[2]{| #1 \rangle\!\langle #2 |}
\newcommand*{\tn}[1]{{\textnormal{#1}}}
\newcommand*{\1}{{\mathbb{1}}}
\newcommand{\T}{\mbox{$\textnormal{Tr}$}}
\newcommand{\todo}[1]{\textcolor[rgb]{0.99,0.1,0.3}{#1}}

\theoremstyle{plain}
\newtheorem{prop}{Proposition}
\newtheorem{proposition}{Proposition}
\newtheorem{theorem}{Theorem}
\newtheorem{lemma}[theorem]{Lemma}
\newtheorem{remark}{Remark}

\theoremstyle{definition}
\newtheorem{definition}{Definition}

\title{Quantum R\'{e}nyi Entropy Functionals for Bosonic Gaussian Systems}
\author{Junseo Lee}
\affiliation{School of Electrical and Electronic Engineering, Yonsei University, Seoul 03722, Korea}
\affiliation{Quantum Computing R\&D, Norma Inc., Seoul 04799, Korea}
\author{Kabgyun Jeong}
\email{kgjeong6@snu.ac.kr}
\affiliation{Research Institute of Mathematics, Seoul National University, Seoul 08826, Korea}
\affiliation{School of Computational Sciences, Korea Institute for Advanced Study, Seoul 02455, Korea}

\date{\today}

\begin{abstract}
In this study, the quantum R\'{e}nyi entropy power inequality of order $p>1$ and power $\kappa$ is introduced as a quantum analog of the classical R\'{e}nyi-$p$ entropy power inequality. To derive this inequality, we first exploit the Wehrl-$p$ entropy power inequality on bosonic Gaussian systems via the mixing operation of quantum convolution, which is a generalized beam-splitter operation. This observation directly provides a quantum R\'{e}nyi-$p$ entropy power inequality over a quasi-probability distribution for $D$-mode bosonic Gaussian regimes. The proposed inequality is expected to be useful for the nontrivial computing of quantum channel capacities, particularly universal upper bounds on bosonic Gaussian quantum channels, and a Gaussian entanglement witness in the case of Gaussian amplifier via squeezing operations.
\end{abstract}

\maketitle

\section{Introduction} \label{intro}
A fundamental open problem addressed in quantum communication theory is understanding the amount of information that can be reliably transmitted through a noisy quantum channel~\cite{BS04,H06}. As a measure of information flow, this quantity is formally characterized by the concept of quantum channel capacity, which is the maximal rate achievable by a quantum communication channel with a vanishing error~\cite{W17,W18}. While the classical capacity of all classical channels is additive, it is known that most quantum channel capacities are surprisingly ``non-additive,'' which indicates that determining the quantum channel capacity is an extremely difficult problem, also referred to as the \emph{superadditivity} of quantum channel capacity~\cite{SY08,H09,LWZG09,LLSSS23}. It is believed that this effect originates from intrinsic quantum phenomena, such as quantum entanglement~\cite{HLW06,HHHH09}.

Although determining the exact channel capacity has not been possible to date, alternative methods have recently been introduced by exploiting a method known as quantum entropy power inequality (EPI)~\cite{KS14,PMG14}, which is a type of entropic functional, providing nontrivial upper bounds on quantum channel capacities, particularly for bosonic Gaussian quantum systems~\cite{KS13,KS13+,JLL19,LLKJ19,J20}. In general, quantum EPI estimates the output entropy of a given quantum channel for independent input signals, and the inequality satisfies a concave property for the functional. In bosonic Gaussian systems, the quantum channel is described by a symplectic unitary transformation, and it is formally given by a beam-splitter or amplifier, which mixes two independent input Gaussian signals into an output signal~\cite{HW01,FOP05,W+12,S17}.

The classical EPI, first introduced by Shannon in 1948~\cite{S48} along with its proofs including extensions~\cite{S59,B65,B75,BL76,L78,DCT91,CS91,VG06,R11,WM14,R17}, can be applied to the context of additive noise channels. Thus, it is still an attractive topic studied in standard information theory to explore the inequality at theoretical levels, such as R\'{e}nyi entropy and the associated functionals~\cite{BC15,BM17}. The basic form of EPI is as follows: For independent random variables $X$ and $Y$ in $\bb{R}^d$ with the corresponding  probability densities $p_X$ and $p_Y$ under a convolution operation $\circledcirc$ such that $(p_X,p_Y)\mapsto p_{X\circledcirc Y}$, the classical EPI is expressed as follows:
\begin{equation*}
\mathbf{V}(X\circledcirc Y)\ge\mathbf{V}(X)+\mathbf{V}(Y),
\end{equation*}
where $\mathbf{V}(X):=\frac{1}{2\pi e}e^{\frac{2}{d}H(X)}$ is the ``entropy power'' or energy, and $H(X):=-\int_{\bb{R}^d}p_X(x)\log p_X(x)\tn{d}x$ is the differential Shannon entropy. The convolution operation on random variables $X$ and $Y$ is defined as $p_{X\circledcirc Y}(z)=\int_{\bb{R}^d} p_X(x)p_Y(z-x)\tn{d}x$ for any $z\in\bb{R}^d$. There are three types of mathematical proofs for EPIs: (i) Gaussian perturbation with de Bruijn's identity and convexity of Fisher information~\cite{S59,B65,CS91,VG06,R11}, (ii) sharp Young's inequality~\cite{B75,BL76,L78,WM14}, and (iii) changes in variables on convex bodies~\cite{R17}. It is known that the exponential form of EPI is essentially equivalent to the linear form, that is, $H(X\circledcirc Y)\ge H(X)+H(Y)$~\cite{DCT91,VG06,R11}. While the classical EPIs are useful for determining the channel capacity such as an additive white Gaussian noise (AWGN) channel, the quantum EPIs are more powerful to quantum Shannon theory when quantum entanglement is involved. Here, the history of quantum EPI, which is a quantum analog of classical EPI, is briefly summarized. The quantum EPI on bosonic Gaussian systems was first derived by K\"{o}nig and Smith~\cite{KS14}, and it was generalized to bosonic Gaussian~\cite{PMG14} and discrete systems~\cite{ADO16}  as well as conditional cases~\cite{K15,JLJ18,PT18,P19}. While the classical capacity is additive,  quantum channel capacities are generally nonadditive; hence, quantum EPIs have the potential power of finding  (tight upper bounds) realistic capacities on quantum channels. As an another potential application, it can be devised to entanglement witness for a continuous-variable (bipartite) quantum system~\cite{DGCZ00,R00,CMYY23} along with a squeezed-type quantum EPIs.

For the classical capacity of a (bosonic) Gaussian quantum channel $\Lambda_{\tau,N_E}$, the nontrivial upper bound is determined by quantum EPI as in~\cite{KS13,KS13+,JLL19}, where $C(\Lambda_{\tau,N_E},N)$ is the classical capacity of the channel, and $\tau$, $N$, and $N_E$ denote the mixing parameter, input, and environment mean photon number, respectively. Then, to use it in an $n$-copy channel, it is known that
\begin{eqnarray*}
C(\Lambda_{\tau,N_E},N)&\le& g(\tau N+(1-\tau)N_E) \\
&&-\lim_{n\to\infty}\min_{\langle\rho_n^G\rangle\le nN}\frac{1}{n}S\left(\Lambda_{\tau,N_E}^{\otimes n}(\rho_n^G)\right),
\end{eqnarray*}
where $\rho_n^G$ denotes the input (entangled) Gaussian quantum state, $g(x):=(x+1)\log(x+1)-x\log x$, and $S(\rho)=-\T\rho\log\rho$ the von Neumann entropy. Also, the minimization is taken over all $n$-mode bosonic Gaussian input states such that its expectation value is less than the total input mean photon number $nN$. By employing the power of quantum EPI, that is, $S\left(\Lambda_{\tau,N_E}^{\otimes n}(\rho_n^G)\right)\ge\tau S(\rho_n^G)+(1-\tau)S({\rho_E^G}^{\otimes n})$, the convex optimization term can be derived as $(1-\tau)g(N_E)$. Note that the lower bound is given by the famous (Gaussian) Holevo capacity formula, and it was reported some additive properties~\cite{GGCH14,LW22} in the Gaussian regime.

The second case is a Gaussian and/or non-Gaussian entanglement witness. In general, a quantum EPI for two-mode Gaussian amplifying channel $\Lambda_\zeta$ can provides an inequality known as 
\begin{equation} \label{eq:gew}
S\left(\Lambda_\zeta(\rho^G)\right)\ge\log(2\zeta-1)
\end{equation}
with a squeezing parameter $\zeta\in(0,1)$, and it can be used to detecting a Gaussian quantum entanglement. (We notice that Eq.~\eqref{eq:gew} is violated, then there exists a Gaussian quantum entanglement.) Because a squeezing-type quantum EPI can be \emph{trivially} recasted as $e^{S\left(\Lambda_\zeta(\rho^G_A)\right)}\ge\max\left\{\zeta e^{S(\rho_A^G)},(\zeta-1)e^{S(\rho_B^G)}\right\}$ from the well-known quantum EPI in the form of $e^{S\left(\Lambda_\zeta(\rho^G_A)\right)}\ge\zeta e^{S(\rho_A^G)}+(\zeta-1)e^{S(\rho_B^G)}$~\cite{PMG14,LLKJ19}. Thus, if we find an entropic measure (or quantity) such as $\mu\left(\Lambda_\zeta(\rho^G)\right)\le\frac{1}{2\zeta-1}$ in the help of the recasted relation above, we can obtain the (non-)Gaussian entanglement witness in Eq.~\eqref{eq:gew}. As an example, $\mu(\varrho):=\T(\varrho^2)$ is a purity of a Gaussian quantum state $\varrho$~\cite{PISS03}.

In this paper, first, a type of quantum R\'{e}nyi EPI on bosonic Gaussian systems is derived through a modification of classical R\'{e}nyi EPIs~\cite{BM17}, with order $p$, power $\kappa$, and weight factor $\tau$. As a concrete example of the bosonic system, first, a (quantum) Wehrl entropy and its associated functional (i.e., Wehrl entropy power) are introduced, including a quantum convolution operation (i.e., $\tau$-beam-splitter), prior to stating the Wehrl EPI. Finally, our main result---the quantum R\'{e}nyi EPI on the bosonic Gaussian system---is formulated. Our results offer a novel perspective on fundamental problems in quantum communication theory in terms of bounding channel capacities. The arguments employed here are expected to be extendable from bosonic systems to arbitrary quantum cases. 

The remainder of this paper is organized as follows. In Section~\ref{sec:REPI}, standard information-theoretic results are provided for the differential R\'{e}nyi EPI, which is further generalized. In Section~\ref{sec:qadd}, a quantum addition rule (in particular, the $\tau$-beam-splitter) is introduced  in the language of a quantum channel. In Section~\ref{sec:WehrlEPI}, a quantum EPI is presented in the framework of Wehrl entropy along with its functional, which are useful in providing our main results. In Section~\ref{sec:QREPI}, the concept of quantum R\'{e}nyi EPI on bosonic Gaussian systems is conceived. Finally, in Section~\ref{sec:conclusion}, a discussion and remarks are presented on this topic, raising some open questions for future research.

\section{R\'{e}nyi entropy power inequality} \label{sec:REPI}
As a variant of the differential Shannon entropy, the differential R\'{e}nyi-$p$ entropy is given by
\begin{equation} \label{eq:renyi-p}
H_p(X)=\frac{1}{1-p}\log\int_{\bb{R}^d}p_X^{~p}(x)\tn{d}x,
\end{equation}
where $X$ is a continuous random variable in $\bb{R}^d$, and $H_1(X)=-\int p_X(x)\log p_X(x)\tn{d}x$ recovers Shannon's variable whenever $p\to1$. It is natural to define R\'{e}nyi-$p$ entropy power as an associated functional of the R\'{e}nyi-$p$ entropy in the form of $\mathbf{V}_p(X)=\exp\left(\frac{2}{d}H_p(X)\right)$, and it is known that, in the sense of a corollary of~\cite{BM17}, the weighted R\'{e}nyi EPI is given by
\begin{equation} \label{eq:cepi-p}
\mathbf{V}_p^\kappa(X\circledcirc_tY)\ge t^\kappa\mathbf{V}_p^\kappa(X)+(1-t)^\kappa\mathbf{V}_p^\kappa(Y),
\end{equation}
where $\circledcirc_t$ denotes the weighted convolution operation with $t\in(0,1)$ and $\kappa\ge\frac{p+1}{2}$ for any $p>1$. 

By exploiting the inequality in Eq.~(\ref{eq:cepi-p}), two types of quantum EPIs, namely Wehrl-$p$ and quantum R\'{e}nyi-$p$ EPIs, are investigated with respect to bosonic Gaussian systems with Husimi and Wigner distributions, respectively. Before the main results, the known ``quantum addition rule'' (or quantum convolution operation) is briefly presented in the context of a quantum channel.

\begin{figure}
\includegraphics[width=8.5cm]{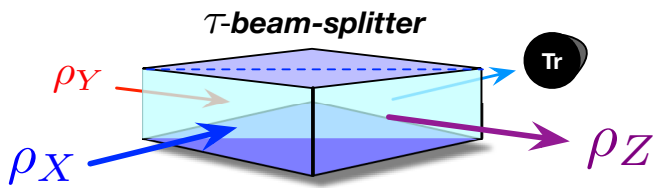}
\caption{Quantum addition rule via $\tau$-beamsplitter operation (or quantum convolution operation) represented by $\rho_Z:=\rho_X\boxplus_\tau\rho_Y$ where $\tau\in(0,1)$}
\label{Fig1}
\end{figure}

\subsection{Quantum addition rule} \label{sec:qadd}
In the classical case, two independent random variables are mixed by applying the convolution operation to the probability densities corresponding to each random variable. However, the quantum case is subtly different. Next, the convolution of quantum states, referred to as the quantum addition rule or \emph{quantum convolution}, is defined~\cite{BGJ23}.

Let $\rho_X$ and $\rho_Y$ be two independent quantum states over $\cl{H}(\bb{C}^d)$, $d$-dimensional complex Hilbert space, and let $\Lambda_\tau$ be a completely positive and trace-preserving (CPT) map satisfying any $\tau\in(0,1)$,
\begin{equation} \label{eq:qadd}
\Lambda_\tau:\rho_X\otimes\rho_Y\mapsto\rho_X\boxplus_\tau\rho_Y=\T_Y\left[U_\tau(\rho_X\otimes\rho_Y)U_\tau^\dag\right],
\end{equation}
where $U_\tau$ denotes any unitary operator over the total quantum system $XY$, $\Lambda_\tau(\bullet)\in\cl{H}(\bb{C}^d)$ is the output of the CPT map, and $\boxplus_\tau$ is the quantum convolution. As an example of quantum convolution, it is possible to define a partial swap operation over $d$-dimensional quantum systems (namely, a qudit-system) with a weight factor $\tau$ as $U_\tau=\sqrt{\tau}\1+i\sqrt{1-\tau}\bb{W}$~\cite{ADO16}, where $\bb{W}$ is the swap operator such that $\bb{W}\rho_{AB}\bb{W}^\dag=\rho_{BA}$. Here, the convolution operation is generalized to the regime of bosonic Gaussian systems (particularly, $\tau$-beam-splitter) and the quantum EPIs are studied.

While the $d$-dimensional quantum states are described on a finite dimensional Hilbert space, bosonic Gaussian states are generally represented in the phase space $\tn{Sp}(2D,\bb{R})$, which is $2D$-dimensional real symplectic space and and it has an infinite dimensional Hilbert space with continuous eigenvalues of Gaussian observables, and evolve with symplectic unitaries ($U_{\tau}^G$) equipped with canonical commutation relation algebra~\cite{FOP05,W+12}. If independent $\rho_X^G$ and $\rho_Y^G$ in symplectic space $\tn{Sp}(2D,\bb{R})$ with mode $D$ are defined as bosonic Gaussian states, the quantum addition rule in Eq.~(\ref{eq:qadd}) can be expressed by the state representation in
$\rho_Z^G:=\rho_X^G\boxplus_\tau\rho_Y^G=\T_Y[U_\tau^G(\rho_X^G\otimes\rho_Y^G)U_\tau^{G^\dag}]$ or the field-operator representation in the form of
\begin{equation} \label{eq:fadd}
\hat{a}_Z=\sqrt{\tau}\hat{a}_X+\sqrt{1-\tau}\hat{a}_Y,
\end{equation}
where $\hat{a}_{\bullet}$ denotes the annihilation operator for the number operator $\hat{n}_{\bullet}:=\hat{a}_{\bullet}^\dag\hat{a}_{\bullet}$ with a creation operator $\hat{a}_{\bullet}^\dag$ on quantum harmonic oscillitors, i.e., $\hat{n}\ket{n}=n\ket{n}$, and $\tau\in(0,1)$ is the mixing parameter (Fig.~\ref{Fig1}).  

Also note that any quantum state of a $D$-mode bosonic system is equivalent to a quasi-probability density function, such as Wigner or Husimi, over the $2D$-dimensional phase space (i.e., real symplectic space). For example, if we introduce  Weyl operator in the form of $\hat{D}({\vec{\eta}}):=\exp\left(i\hat{x}^T\Gamma\vec{\eta}\right)$ for $\vec{\eta}\in\bb{R}^{2D}$, then a density operator $\rho^G$ is represented by the Wigner characteristic function and Wigner function (via Fourier transform) as follows:
\begin{align*}
\chi(\vec{\eta})&=\T\left(\rho^GD(\vec{\eta})\right)\;\;\tn{and} \\
W(\hat{x})&=\frac{1}{(2\pi)^{2D}}\int_{\bb{R}^{2D}}\exp\left(-i\hat{x}^T\Gamma\vec{\eta}\right)\chi(\vec{\eta})\tn{d}^{2D}\vec{\eta},
\end{align*}
respectively. Here, $\hat{x}=(\hat{q}_1,\hat{p}_1,\ldots,\hat{q}_D,\hat{p}_D)^T$ with a commutation relation $[\hat{x}_j,\hat{x}_k]=2i\Gamma_{jk}$ denotes the quadrature field operator, and $\Gamma:=\bigoplus_{j=1}^D\gamma$ with
\begin{equation*}
\gamma= \begin{pmatrix}
0 & 1 \\
1 & 0 \\ 
\end{pmatrix}.
\end{equation*}
Thus, any density matrix $\rho^G$ can be equivalently described by a quasi-probability distribution (i.e., it has a normalization factor 1 but generally non-positive)~\cite{FOP05,W+12}.

In addition, the R\'{e}nyi entropy of order $p$ of a $d$-dimensional quantum state $\rho\in\cl{H}(\bb{C}^d)$ can be formally defined as
\begin{equation}
S_p(\rho)=\frac{1}{1-p}\log\T\rho^p
\end{equation}
for any $p>0~(p\neq1)$, and converges to the von Neumann entropy in the limit of $p\to1$. However, a differential version of R\'{e}nyi entropy is required and thus, the related entropic functional on the bosonic Gaussian system is explained in detail in Section~\ref{sec:QREPI}. First, consider the Wehrl entropy and its functional.

\section{Wehrl entropy and associated entropy power inequality} \label{sec:WehrlEPI}
The Wehrl entropy and  associated entropy power of order $p$ on a bosonic Gaussian state $\rho_X\in\tn{Sp}(2D,\bb{R})$ can be defined as
\begin{eqnarray}
S_p^W(\rho_X)&=&\frac{1}{1-p}\log\int_{\bb{C}^D}Q_X^{~p}(\alpha)\tn{d}\alpha \label{eq:wehrl}\\
\mathbf{V}_p^W(\rho_X)&=&\exp\left(\frac{1}{D}S_p^W(\rho_X)\right), \label{eq:wep}
\end{eqnarray}
where $Q_X$ denotes the Husimi distribution corresponding to quantum state $\rho_X$, which is formally given by $Q_X(\alpha)=\frac{1}{(2\pi)^D}\bra{\alpha}\rho_X\ket{\alpha}$ with coherent state $\ket{\alpha}=e^{-|\alpha|^2/2}\sum_{n=0}^\infty\frac{\alpha^n}{\sqrt{n!}}\ket{n}~(\forall\alpha\in\bb{C})$; moreover, note that $\int_{\bb{C}^D}Q_X(\alpha)\tn{d}\alpha=1$ and $Q_X(\alpha)\ge0$.
For this distribution, the quantum convolution with a mixing parameter $\tau\in(0,1)$ is given by~\cite{KI95}
\begin{equation} \label{eq:Hconv}
Q_{Z}(\beta)=\frac{1}{\tau}\int_{\bb{C}^D}Q_X(\alpha)Q_Y\left(\frac{\beta-\sqrt{1-\tau}\alpha}{\sqrt{\tau}}\right)\tn{d}\alpha,
\end{equation}
which corresponds to $\rho_Z^G=\rho_X^G\boxplus_\tau\rho_Y^G$ or Eq.~(\ref{eq:fadd}) for a bosonic Gaussian system involving the Husimi distribution $Q$.

The formulas for differential Shannon entropy and Wehrl entropy have similar forms in the logarithmic components; thus, the proof techniques introduced in the literature can be applied in a straightforward manner to classical R\'{e}nyi EPIs~\cite{J20,BM17}---modified Young's inequality in the power $\kappa$. Then, the Wehrl EPI of order $p$ is obtained as follows:

\begin{theorem}[Wehrl-$p$ EPI] \label{thm1}
Let $\rho_X^G$ and $\rho_Y^G$ in $\tn{Sp}(2D,\bb{R})$ be any independent $D$-mode Gaussian states corresponding to Husimi distributions $Q_X$ and $Q_Y$, respectively. Let $\tau\in(0,1)$ be the mixing parameter. Then,
\begin{eqnarray} \label{eq:wepi-p}
\mathbf{V}_p^{W,\kappa}(\rho_Z^G)\ge\tau^\kappa\mathbf{V}_p^{W,\kappa}(\rho_X^G)+(1-\tau)^\kappa\mathbf{V}_p^{W,\kappa}(\rho_Y^G),
\end{eqnarray}
where $\rho_Z^G=\rho_X^G\boxplus_\tau\rho_Y^G$ is the output bosonic Gaussian state
 with power $\kappa\ge\frac{p+1}{2}$ for any $p>1$.
\end{theorem}

The explicit proof of {\bf Theorem~\ref{thm1}} is provided in Appendix A and Appendix B. It is natural that, as a generalization of Wehrl-$p$ entropy and EPI (i.e., a probability distribution), a ``quantum R\'{e}nyi entropy'' and its associated entropic functional (i.e., a quasi-probability distribution) be considered on the bosonic Gaussian systems. To address a general case in Gaussian regime, the Husimi distribution $Q_X$ can be replaced with a quasi-probability distribution density function $q_X$ (e.g., Wigner) corresponding to the bosonic system $\rho_X$. In general, the Husimi distribution function has only positive values, but Wigner function can be negative in the probability space. Moreover, it is well-known that the Husimi function is well-behaved and non-singular for all Gaussian states, whereas generic quasi-probability functions can potentially exhibit a singularity or ill-defined behavior even for Gaussian states. In this reason, it is believed that a R\'{e}nyi-$p$ EPI with a Wigner function allows us a strong bound instead of the Wehrl-$p$ EPI with a Husimi function, which imply some potential applications such a Gaussian entanglement detection or determination of Gaussian channel capacities.  Here, a generalized quantum convolution of two independent quasi-distributions takes the form $q_Z(z')=\frac{1}{\tau}\int_{\bb{C}^D} q_X(z)q_Y(\frac{z'-\sqrt{1-\tau}z}{\sqrt{\tau}})\tn{d}z$, where $z,z'\in{\bb{C}}^D$, as in Eq.~(\ref{eq:Hconv}). The main result can now be stated in the following.

\begin{table*}
\caption{Summary of R\'{e}nyi-$p$ entropies and its functionals.}
\begin{ruledtabular}
\begin{tabular}{l c c c}
{} &  Classical R\'{e}nyi  & Wehrl (quantum)  & Quantum R\'{e}nyi  \\
\hline
{\bf State rep. }  &  $X,Y\in\bb{R}^d$ & $\rho_X,\rho_Y\in\tn{Sp}(2D,\bb{R})$ & $\rho_X,\rho_Y\in\tn{Sp}(2D,\bb{R})$ \\
{\bf PDFs}  &  $p_X,p_Y$ & $Q_X,Q_Y$ & $q_X,q_Y$ \\
{\bf Entropy} &   $H_p(X)=\frac{1}{1-p}\log\int_{\bb{R}^d}p_X^p(x)\tn{d}x$ & $S_p^W(\rho_X)=\frac{1}{1-p}\log\int_{\bb{C}^{D}}Q_X^p(\alpha)\tn{d}\alpha$ & $S_p(\rho_X)=\frac{1}{1-p}\log\int_{\bb{C}^{D}}q_X^p(z)\tn{d}z$ \\
{\bf Parameter}  &   $t\in(0,1)$ & $\tau\in(0,1)$ & $\tau\in(0,1)$ \\
{\bf Addition} &   $Z=X\circledcirc_tY=\sqrt{t}X+\sqrt{\tilde{t}}Y$ & $\rho_Z=\rho_X\boxplus_\tau\rho_Y$ & $\rho_Z=\rho_X\boxplus_\tau\rho_Y$ \\
{\bf $L^p$-norm} &  $\|p_X\|_p=\left(\int_{\bb{R}^d}p_X^p(x)\tn{d}x\right)^{1/p}$ & $\|Q_X\|_p=\left(\int_{\bb{C}^D}Q_X^p(\alpha)\tn{d}\alpha\right)^{1/p}$ & $\|q_X\|_p=\left(\int_{\bb{C}^D}q_X^p(z)\tn{d}z\right)^{1/p}$ \\
{\bf Mixing ops. }  & $p_Z(z)=\int_{\bb{R}^d}p_X(x)p_Y(z-x)\tn{d}x$ & $Q_Z(\beta)=\frac{1}{\tau}\int_{\bb{C}^D}Q_X(\alpha)Q_Y(\frac{\beta-\sqrt{\tilde{\tau}}\alpha}{\sqrt{\tau}})\tn{d}\alpha$ & $q_Z(z')=\frac{1}{\tau}\int_{\bb{C}^D}q_X(z)q_Y(\frac{z'-\sqrt{\tilde{\tau}}z}{\sqrt{\tau}})\tn{d}z$ \\
{\bf Power}  & $\kappa\ge\frac{p+1}{2}~(p>1)$ & $\kappa\ge\frac{p+1}{2}~(p>1)$ & $\kappa\ge\frac{p+1}{2}~(p>1)$ \\
{\bf EPIs}  &   $\mathbf{V}_p^\kappa(Z)\ge t^{\kappa}\mathbf{V}_p^\kappa(X)+\tilde{t}^\kappa \mathbf{V}_p^\kappa(Y)$ & $\mathbf{V}_p^{W,\kappa}(\rho_Z)\ge \tau^\kappa \mathbf{V}_p^{W,\kappa}(\rho_X)+\tilde{\tau}^\kappa \mathbf{V}_p^{W,\kappa}(\rho_Y)$ & $\mathbf{V}_p^\kappa(\rho_Z)\ge \tau^\kappa \mathbf{V}_p^\kappa(\rho_X)+\tilde{\tau}^\kappa \mathbf{V}_p^\kappa(\rho_Y)$ \\
\end{tabular}
\end{ruledtabular}
\label{table:epi}
Notice that $\circledcirc$:~convolution, $t,\tau\in(0,1)$: mixing parameters, $\boxplus_\tau$:~$\tau$-beamsplitter, $D$:~$D$-mode, $d$:~$d$-dimension, $\tilde{t}=1-t$,  $\tilde{\tau}=1-\tau$.
\end{table*}

\section{Quantum R\'{e}nyi entropy power inequality} \label{sec:QREPI}
In the framework of arbitrary $D$-mode bosonic Gaussian systems, quantum (differential) R\'{e}nyi-$p$ entropy, as an extension of Eq.~(\ref{eq:wehrl}), can be defined as
\begin{equation}
S_p(\rho_X)=\frac{1}{1-p}\log\int_{\bb{C}^{D}}q_X^p(z)\tn{d}z
\end{equation}
for any order $p>0~(p\neq1)$ and for any quasi-probability distribution $q_X$ corresponding to $\rho_X\in\tn{Sp}(2D,\bb{R})$. From this, it is also natural to define an associated functional, namely quantum R\'{e}nyi-$p$ entropy power, as $\mathbf{V}_p(\rho_X)=\exp\left(\frac{1}{D}S_p(\rho_X)\right)$ for the bosonic Gaussian state. Adopting the logarithmic form and without loss of generality, the sharp Young's inequality in~\cite{BM17} can be used again.

In mixing input signals, the beam-splitter operation, $\boxplus_\tau$, is referred to as quantum convolution, which can be obtained by substituting the Husimi $Q$-function with an arbitrary probability ($q_{\bullet}$) on bosonic Gaussian states. The field operator representation is described in the form of Eq.~(\ref{eq:fadd}), with a weight factor of $\tau\in(0,1)$.

\begin{theorem}[Quantum R\'{e}nyi-$p$ EPI] \label{thm:main}
For any $D$-mode Gaussian states $\rho_X^G$ and $\rho_Y^G$ in $\tn{Sp}(2D,\bb{R})$, and for any $\tau\in(0,1)$, we have
\begin{equation} \label{eq:qrepi-p}
\mathbf{V}_p^\kappa(\rho_X^G\boxplus_\tau\rho_Y^G)\ge \tau^\kappa\mathbf{V}_p^\kappa(\rho_X^G)+(1-\tau)^\kappa\mathbf{V}_p^\kappa(\rho_Y^G),
\end{equation}
where the power satisfies $\kappa\ge\frac{p+1}{2}$ for any $p>1$.
\end{theorem}

More precisely, it can be expressed using the exponential form as
\begin{eqnarray*}
&&\exp\left(\frac{\kappa}{D}S_p(\rho_X\boxplus_\tau\rho_Y)\right) \\
&&\ge\tau^\kappa\exp\left(\frac{\kappa}{D}S_p(\rho_X)\right)+(1-\tau)^\kappa\exp\left(\frac{\kappa}{D}S_p(\rho_Y)\right).
\end{eqnarray*}
The components of EPIs in both classical and quantum cases are summarized in Table~\ref{table:epi}. Furthermore, note that the inequality in Eq.~(\ref{eq:qrepi-p}) can be easily derived from the fundamental quantum EPI for $\kappa=1$~\cite{KS14,PMG14}: $\mathbf{V}_p^1(\rho_Z)\ge \tau\mathbf{V}_p^1(\rho_X)+(1-\tau)\mathbf{V}_p^1(\rho_Y)$ because the $\kappa$-power of the total equation and removing trivial (positive) terms on the right-hand side generates our main claim in Theorem~\ref{thm:main}. However, there are some drawbacks to $\kappa>1$ in classical as well as quantum R\'{e}nyi EPIs when considering Young's constant argument.

\section{Conclusions} \label{sec:conclusion}
The quantum entropy power inequalities are very powerful tool for determining quantum channel capacities in quantum Shannon theory when quantum entanglement is involved.
In this paper, an identity called quantum R\'{e}nyi EPI of order $p>1$ and power $\kappa$ is obtained via a generalization of Wehrl-$p$ EPI on bosonic Gaussian regimes based on the classical R\'{e}nyi-$p$ EPIs, which can be useful for the nontrivial computing of quantum channel capacities. In the quantum case, it is conjectured that $\mathbf{V}_p^1(\rho_X\boxplus_\tau\rho_Y)\ge \tau\mathbf{V}_p^1(\rho_X)+(1-\tau)\mathbf{V}_p^1(\rho_Y)$ for the order $p>0$; however, this still requires an explicit proof. The main result (i.e., Theorem~\ref{thm:main}) is true for any order of $p$ in the R\'{e}nyi functionals. In particular, it can help achieve a recent result of the modified breakthrough of Young's convolution inequality.

In the sense of explicit computing of the quantum channel capacities, this type of research is the only way to highlight open problems, but it is still far from reaching the exact solution. As an example, it is expected that the following nontrivial upper bound of the channel capacity on $\Lambda_{\tau,N_E}$ can be computed. That is, for all $p>0$ and all $\rho^G\in\tn{Sp}(2D,\bb{R})$,
\begin{eqnarray*}
&&C(\Lambda_{\tau,N_E},N) \\
&& \le\max_{\rho^G}S_p\left(\Lambda_{\tau,N_E}(\rho^G)\right)-\lim_{n\to\infty}\min_{\langle\rho_n^G\rangle\le nN}\frac{1}{n}S_p\left(\Lambda_{\tau,N_E}^{\otimes n}(\rho_n^G)\right),
\end{eqnarray*}
where parameters $\tau$, $\kappa$, and $p$ encompass all ranges in terms of an extended notion of quantum EPIs. Furthermore, we remark that, for any Gaussian amplifying channel with a squeezing parameter $\zeta$, it might be possible to construct an efficient Gaussian entanglement measure such that $S_p\left(\Lambda_\zeta(\rho^G)\right)<\log(2\zeta-1)$ under the ordering of $S(\rho^G)\ge\cdots\ge S_p(\rho^G)$ for a symmetric Gaussian state as well as non-Gaussian cases. In addition, neither the nonadditive property of qudit systems nor the subtle major problem of the entropy photon number inequality (EPNI) in the Gaussian regime are known. EPNI was conjectured by Guha \emph{et al.}~\cite{GES08} and is given by $\mathbf{N}(\rho_X\boxplus_\tau\rho_Y)\ge\tau\mathbf{N}(\rho_X)+(1-\tau)\mathbf{N}(\rho_Y)$, where $\mathbf{N}(\rho)=g^{-1}(\frac{S(\rho)}{D})$, which is the ``photon number'' of $D$-mode $\rho$. If this conjecture is true,  the classical capacities of various bosonic Gaussian quantum channels can be evaluated~\cite{GSE07}.

Because most advances in quantum Shannon theory emerge from information theory, there are still reservations regarding the EPI itself---ranging from the classical, Tsallis, and unified EPIs to the unknown quantum perspective. Furthermore, there are several intriguing open questions on the quantum EPI. First, it is not known whether other types of nontrivial quantum convolutions on qudit systems beyond the usual beam-splitter or amplifier on bosonic quantum systems exist. Second, it is not known how far the resulting quantum EPIs are from known lower bounds, such as the Holevo capacity, data-processing-inequalities, or coherent and private information. This kind of research may extend our knowledge of original quantum channel capacity problems while providing novel insights into the theory and applications of emerging quantum networks or quantum internet.
\bigskip

\section*{Acknowledgments}
The authors thank to anonymous reviewer's valuable comments. This work was supported by the National Research Foundation of Korea (NRF) through a grant funded by the Ministry of Science and ICT (NRF-2020M3E4A1077861, NRF-2022M3H3A1098237) and the Ministry of Education (NRF-2021R1I1A1A01042199). This research was partly supported by Korea Institute of Science and Technology Information (KISTI, P23031).

%


\section*{Appendix}

\bigskip
\begin{center}
{\large\bf Quantum R\'{e}nyi Entropy Functionals for Bosonic Gaussian Systems} \\
\bigskip
{\large Kabgyun Jeong \& Junseo Lee}
\end{center}
\bigskip

\subsection{Sharp Young's inequality}\label{appendix:A}

Some mathematical background is required for the proof of {\bf Theorem 1}, which is essentially based on the modification of Bobkov and Marsiglietti's derivation~\cite{BM17} for the classical R\'{e}nyi entropy power inequality. Notably, the Husimi distribution of a Gaussian quantum state is almost equivalent to that of a classical density function.

First, the $L^p$-norm for the positive Husimi function $Q_X$ in the Gaussian regime is defined as follows:
\begin{equation}
\|Q_X\|_p=\left(\int_{\bb{C}^D}Q_X^p(\alpha)\tn{d}\alpha\right)^{1/p}.
\end{equation}

Then, let $\rho_X^G$ and $\rho_Y^G\in\tn{Sp}(2D,\bb{R})$ be independent Gaussian quantum states associated with Husimi functions $Q_X$ and $Q_Y$, respectively. The sharp Young's inequality implies that, for any $r$, $s$, and $t\ge1$ such that $\frac{1}{r}+\frac{1}{s}-\frac{1}{t}=1$,
\begin{equation}
\|Q_X\circledcirc_\tau Q_Y\|_t\le K^{D/2}\|Q_X\|_r\|Q_Y\|_s,
\end{equation}
where $K=\frac{k_rk_s}{k_t}\le1$ with $k_p=\left(1-\frac{1}{p}\right)\sqrt[\leftroot{0} \uproot{5}p]{\frac{p^2}{p-1}}$. Note that $\circledcirc_\tau$ denotes the quantum convolution (or $\tau$-beam-splitter), as in Eq.~(\ref{eq:Hconv}). In addition, any contribution of mixing parameter $\tau$ can be resolved at the end of the proof of {\bf Theorem 1} as a natural generalization of Eq.~(\ref{eq:cepi-p}).

By the way, the relationship between $L^p$-norm and R\'{e}nyi-$p$ entropy is easily induced as follows:
\begin{eqnarray}
\|Q_X\|_p \nonumber
&=&\left(\int_{\bb{C}^D}Q_X^p(\alpha)\tn{d}\alpha\right)^{1/p}=\exp\left(\frac{1}{p}\log\int_{\bb{C}^D}Q_X^{~p}(\alpha)\tn{d}\alpha\right) \\ \nonumber
&=&\exp\left(\frac{1-p}{p}S_p(\rho_X)\right) =\mathbf{V}_p(\rho_X^G)^{-\frac{D}{2}\left(1-\frac{1}{p}\right)}.
\end{eqnarray}

By modifying Equation (A.2) through this property, the following can be obtained.
\begin{equation} \label{eq:rst1}
\mathbf{V}_t(\rho_X^G\boxplus_\tau\rho_Y^G)^{1-\frac{1}{t}}\ge\frac{1}{K}\mathbf{V}_r(\rho_X^G)^{1-\frac{1}{r}}\mathbf{V}_s(\rho_Y^G)^{1-\frac{1}{s}}.
\end{equation}
As a quantum analog, this result corresponds to Beckner's information-theoretic formulation~\cite{B75}.

By exploiting H\"{o}lder's inequality, Eq.~(\ref{eq:rst1}) can be substituted into one functional $\mathbf{V}_t(\cdot)$ as follows: For any $r,s,t\ge 1$ satisfying the constant $K$,
\begin{equation} \label{eq:rst2}
\mathbf{V}_t(\rho_X^G\boxplus_\tau\rho_Y^G)^{1-\frac{1}{t}}\ge\frac{1}{K}\mathbf{V}_t(\rho_X^G)^{1-\frac{1}{r}}\mathbf{V}_t(\rho_Y^G)^{1-\frac{1}{s}}.
\end{equation}
To prove the main result, two technical lemmas (Sec. IV and its derivations in~\cite{BM17}) are introduced. \\

\textbf{Lemma 1} (Homogeneity property).
For any $t>1$, suppose that any positive $a$ and $b$ satisfy $a+b=1-\frac{1}{t}$. Then, $r,s\ge1$ exists such that
\begin{equation}
K^{-\frac{\kappa t}{t-1}}a^{\frac{t(r-1)}{r(t-1)}}b^{\frac{t(s-1)}{s(t-1)}}\ge1-\frac{1}{t},
\label{eq:hom}
\end{equation}
where the constant $K=\frac{k_rk_s}{k_t}$. \\

\textbf{Lemma 2} (Calculus lemma).
For a given $0<c<1$ and $d\ge\frac{2}{c}-1$, the algebraic function is
\begin{equation}
F(x)=\frac{(1-x)^{d(1-x)}(1-y)^{d(1-y)}}{x^xy^y}\;\;\;(y=c-x)
\label{eq:cal}
\end{equation}
attaining a minimum in the region of $0\le x \le c$ either at the points $x=0$, $x=c$, or at the center $x=\frac{c}{2}$. Furthermore, in the case of $d=\frac{2}{c}-1$, the function reaches its minimum at the endpoints.

\subsection{Proof of Theorem 1}\label{appendix:B}

{\it Proof}. First, recall the inequality given in Eq.~(\ref{eq:rst2}). Now, let us rewrite the inequality by using $\kappa=\frac{t+1}{2}$ in \textbf{Lemma 1} such that
\begin{equation}
\mathbf{V}_t^\kappa(\rho_X^G\boxplus_\tau\rho_Y^G)^{1-\frac{1}{t}}\ge\frac{1}{K^{\frac{\kappa t}{t-1}}}\mathbf{V}_t^\kappa(\rho_X^G)^{\frac{t(r-1)}{r(t-1)}}\mathbf{V}_t^\kappa(\rho_Y^G)^{\frac{t(s-1)}{s(t-1)}}.
\end{equation}

Finally, by substituting $a=\mathbf{V}_t^\kappa(\rho_X^G)$ and $b=\mathbf{V}_t^\kappa(\rho_Y^G)$ and exploiting the assumption of $a+b=1-\frac{1}{t}$ in {\bf Lemma~1}, we have
\begin{equation} \label{eq:fin}
K^{-\frac{\kappa t}{t-1}}a^{\frac{t(r-1)}{r(t-1)}}b^{\frac{t(s-1)}{s(t-1)}}\ge1-\frac{1}{t}
\end{equation}
for some admissible $r,s\ge1$ such that $\frac{1}{r}+\frac{1}{s}-\frac{1}{t}=1$. This completes the proof. $\blacksquare$

\bigskip
Note that Eq.~(\ref{eq:fin}) implies that 
\begin{eqnarray*}
\mathbf{V}_t^{W,\kappa}(\rho_X^G\boxplus_\tau\rho_Y^G)\ge\tau^\kappa\mathbf{V}_t^{W,\kappa}(\rho_X^G)+(1-\tau)^\kappa\mathbf{V}_t^{W,\kappa}(\rho_Y^G)
\end{eqnarray*}
is true, and the quantum generalization of the classical case given in Eq.~(\ref{eq:cepi-p}) can be utilized.

\end{document}